\newcommand{\be}{\begin{equation}}\newcommand{\ee}{\end{equation}}
\newcommand{\bea}{\begin{eqnarray}}\newcommand{\eea}{\end{eqnarray}}
\newcommand{\nn}{\nonumber\\[6pt]}
\newcommand{\p}[1]{(\ref{#1})}

\newcommand{\bD}{\overline D}
\newcommand{\cD}{{\cal D}}
\newcommand{\cW}{{\cal W}}
\newcommand{\cbD}{{\overline{\cal D}}}
\newcommand{\bT}{{\overline T}}
\newcommand{\bV}{{\overline V}}
\newcommand{\bQ}{{\overline Q}}
\newcommand{\bS}{{\overline S}}
\newcommand{\bt}{{\bar\theta}}

\newcommand{\bpsi}{{\bar\psi}}
\newcommand{\brho}{{\bar\rho}}
\newcommand{\bphi}{{\bar\phi}}
\newcommand{\blam}{{\bar\lambda}}
\newcommand{\bLam}{{\overline\Lambda}}
\newcommand{\bomega}{{\bar\omega}}
\newcommand{\beps}{{\bar\epsilon}}
\newcommand{\vt}{\vartheta}
\newcommand{\vf}{\varphi}
\newcommand{\br}{{\bf r}}
\newcommand{\bL}{{\bf L}}
\newcommand{\bJ}{{\bf J}}
\newcommand{\bI}{{\bf I}}
\newcommand{\eps}{\epsilon}
\newcommand{\sfrac}[2]{{\textstyle\frac{#1}{#2}}}

\documentclass[12pt]{article}
\usepackage{amscd,amsmath,amssymb}

\begin{document}

\thispagestyle{empty}
\vspace{2cm}
\begin{flushright}
hep-th/0212303 \\
ITP-UH-25/02 \\[5mm]
December, 2002\\[3cm]
\end{flushright}
\begin{center}
{\Large\bf New variant of N=4 superconformal mechanics }
\end{center}
\vspace{1cm}

\begin{center}
{\large\bf  E. Ivanov${}^{a}$,
S. Krivonos${}^{a}$, O. Lechtenfeld${}^{b}$ }
\end{center}

\begin{center}
${}^a$ {\it Bogoliubov  Laboratory of Theoretical Physics, JINR, 141980 Dubna,
Russia}

\vspace{0.2cm}
${}^b$ {\it Institut f\"ur Theoretische Physik, Universit\"at Hannover,} \\
{\it Appelstra\ss{}e 2, 30167 Hannover, Germany}
\end{center}
\vspace{2cm}

\begin{abstract}
\noindent
Proceeding from a nonlinear realization of the most general $N{=}4, d{=}1$
superconformal symmetry, associated with the supergroup $D(2,1;\alpha)$,
we construct a new model of nonrelativistic $N{=}4$ superconformal mechanics.
In the bosonic sector it combines the worldline dilaton with the fields
parametrizing the $R$-symmetry coset $S^2 \sim SU(2)/U(1)$. We present
invariant off-shell $N{=}4$ and $N{=}2$ superfield actions for this system
and show the existence of an independent $N{=}4$ superconformal invariant
which extends the dilaton potential. The extended supersymmetry requires
this potential to be accompanied by a $d{=}1$ WZW term on $S^2$. We study
the classical dynamics of the bosonic action and the geometry of its
sigma-model part. It turns out that the relevant target space is a cone
over $S^2$ for any non-zero $\alpha\neq\pm\frac12$. The constructed model
is expected to be related to the `relativistic' $N{=}4$ mechanics of
the AdS$_2\times S^2$ superparticle via a nonlinear transformation
of the fields and the time variable.

\end{abstract}

\newpage
\setcounter{page}{1}
\section{Introduction}
Models of conformal and superconformal quantum mechanics \cite{dff}-\cite{stro2}
are important basically due to their role as `conformal field theories'
in the AdS$_2$/CFT$_1$ version of the AdS/CFT correspondence \cite{ads},
the fact that they describe the near-horizon dynamics
of the black-hole solutions of supergravity, and their tight relation
with integrable Calogero-Moser type systems.

The recent renewal of interest in (super)conformal mechanics models
was mainly triggered by the observation \cite{nscm} that
the radial motion of a superparticle at  zero angular momentum in
the near-horizon region of the extreme Reissner-Nordstr\"om black hole is 
governed by a modified (`new' or `relativistic') version of conformal mechanics.
The corresponding background is AdS$_2\times S^2$,
with the radial coordinate of the superparticle parametrizing its AdS$_2$ part.
Since the full isometry of this background is the supergroup $SU(1,1\vert 2)$,
it was suggested in \cite{nscm} that the full (radial and angular) dynamics
of this superparticle is described by a `new' variant of $N{=}4$ superconformal
mechanics~\footnote{
By $N$ we always count {\it real} $d{=}1$ Poincar\'e supercharges.
This convention differs from the nomenclature due to ref.~\cite{wit1} in
which $N$ denotes the number of {\it complex} $d{=}1$ Poincar\'e supercharges.}
containing three physical bosonic fields in its supermultiplet
(the radial AdS$_2$ coordinate and two angular coordinates parametrizing $S^2$).
Yet, even the `old' conformal mechanics~\cite{dff} has not been extended in 
full generality to an $N{=}4$ superconformal mechanics 
(with additional angular fields).
The $N{=}4$ superconformal mechanics constructed in \cite{ikl2}
contains only one bosonic physical field in its off-shell supermultiplet
(namely the dilaton associated with the generator of spontaneously broken
dilatations). In this sense the model represents a minimal $N{=}4$ extension
of the standard one-component conformal mechanics of ref.\cite{dff}.
An attempt to define the most general superconformal mechanics
on $SU(1,1\vert 2)$ using the nonlinear realization approach was undertaken
in \cite{7}. However, neither an invariant action nor the solution of the
irreducibility constraints on the Goldstone superfields were presented.

A Green-Schwarz-type action for the AdS$_2\times S^2$ superparticle was
constructed in \cite{part}. After properly gauge-fixing kappa-symmetry and
worldline diffeomorphisms the corresponding action can be treated as an $N{=}4$
superconformal mechanics action extending that of the `new' conformal mechanics.
However, the $N{=}4$ superconformal symmetry in such an action is non-manifest
and closes only on shell, which is typical for the gauge-fixed
Green-Schwarz-type actions.

There are two basic ways of constructing superconformally invariant actions in
one dimension. One of them proceeds from the most general $N{=}1$ supersymmetric
sigma model on the $d{=}1$ worldline of the particle (in general 
supplemented by a coupling to an external electromagnetic potential). 
Then one looks for the restrictions the extra
superconformal invariance imposes on the geometry of such a sigma model
(see e.g. \cite{papa1}-\cite{stro2}). In this approach one as a rule
deals with component fields (or at most $N{=}1$ superfields), 
and the extra Poincar\'e and conformal supersymmetries are not manifest. 
The other approach, pioneered in \cite{ikl2} and later employed in \cite{7},
starts from nonlinear realizations of appropriate superconformal symmetries in
terms of a set of worldline Goldstone superfields. 
It is a natural generalization of the approach of ref. \cite{ikl1} 
where conformal mechanics \cite{dff}
was shown to have a natural description in terms of a nonlinear realization
of $SO(1,2)$, the conformal group in one dimension. In this way
the aforementioned minimal $N{=}4$ superconformal mechanics,
associated with the supergroup $SU(1,1\vert2)$ and having only a dilaton 
field in the bosonic sector, was found and studied~\cite{ikl2}.
Besides keeping manifest and off-shell the complete set of $d{=}1$ Poincar\'e
supersymmetries, this second approach has the advantage that it automatically
generates the correct irreducibility conditions for the basic Goldstone
superfields and ensures a geometric interpretation for their components
(the bosonic and fermionic $d{=}1$ Goldstone fields) as parameters of the
relevant supercosets.

In the present paper,
following the nonlinear realization approach of \cite{ikl2}, we construct
the most general $N{=}4$ superfield extension of the `old' conformal mechanics
with additional angular fields in the bosonic sector. We start with the
exceptional supergroup $D(2,1;\alpha)$ \cite{FRS} as the most general $N{=}4$
superconformal group in one dimension. After the appropriate choice of the 
supercoset, the only Goldstone $N{=}4$ superfields surviving the inverse Higgs
procedure \cite{IH} are the superdilaton and the two parameters of the coset
$SU(2)/U(1)$ (a second $SU(2) \subset D(2,1;\alpha)$ is placed into the
stability subgroup and acts on the fermions only). For these three superfields
there naturally appear simple irreducibility conditions.
Surprisingly, the latter prove to be identical to those used in \cite{ismi}
for constructing the off-shell $N{=}4$ supersymmetric $1d$ sigma model action
with a bosonic target manifold of dimension $3k$. It is easy to solve these
constraints through an unconstrained prepotential and to construct two
independent superfield invariants which in components give rise to the kinetic 
term and a scalar potential term. The latter is necessarily accompanied by a 
worldline WZW term on $S^2$ which is nothing but the coupling of the charged
particle to a Dirac magnetic monopole (see e.g. \cite{plyu}). Besides the $N=4$
superfield form of the invariant action we present its $N=2$ superfield form.
As was recently shown \cite{ikn}, the `old' and `new' conformal mechanics
models are actually related by a nonlinear transformation of the time
variable and the coordinate fields (it is a particular case of the `holographic'
transformation introduced in \cite{bik3}). We expect that this equivalence
extends to the case we are dealing with. Namely, at least for the special case
of $D(2,1;\alpha{=}{-}1)\sim SU(1,1\vert2)\subset\!\!\!\!\!\times\,SU(2)$ 
we surmise the
existence of a nonlinear transformation which maps the $N{=}4$ superconformal
mechanics we shall presently construct to the `new' $N{=}4$ superconformal
mechanics of the AdS$_2\times S^2$ superparticle. On shell its action should
coincide with the gauge-fixed action of ref.~\cite{part}.\footnote{
At the on-shell Hamiltonian level, the issue of canonical equivalence between
some versions of these two, at first sight different, N{=}4 superconformal 
mechanics systems is studied in \cite{new}.}

\section{D(2,1;$\alpha$) supergroup as N=4 superconformal group in d=1
and its nonlinear realization}
We use the standard definition  of the superalgebra $D(2,1;\alpha)$ \cite{FRS} 
with the notations of ref. \cite{Iv1}. It contains the following generators.

Spinor generators:
\be\label{def1}
Q^{Ai'i}\;,\quad \overline{\left( Q^{Ai'i}\right)}=
  \epsilon_{ik}\epsilon_{i'k'}Q^{Ak'k} \; , \quad (A,i,i'=1,2) \; .
\ee

Bosonic generators:
\be\label{def1a}
T_2^{AB}\; , \quad T^{i'j'} \; , \quad T_1^{ij}~.
\ee

The indices $A, i$ and $i'$ refer to fundamental representations of
the mutually commuting $sl(2,R)$ and two $su(2)$ algebras which form the bosonic
sector of $D(2,1;\alpha)$.  The bosonic generators obey the following conjugation rules:
\be
\left( T_2^{AB}\right)^\dagger =  T_2^{AB}\; , \quad
\left( T^{11}\right)^\dagger =  T^{22}\; , \;
\left( T^{12}\right)^\dagger =  -T^{12}\; ,\quad
\left( T_1^{11}\right)^\dagger =  T_1^{22}\; , \;
\left( T_1^{12}\right)^\dagger =  -T_1^{12}\;,
\ee
and satisfy the commutation relations (common for all $T$):
\be
\left[ T^{ab}, T^{cd}\right]= -i\left( \epsilon^{ac}T^{bd}+
 \epsilon^{bd}T^{ac} \right),
\ee
where $a,b$ stand for any sort of doublet indices ($A, i$ or $i'$).
The commutator of any generator $T^a$ with $Q^b$ (other indices of $Q^{Aii'}$
being suppressed) reads:
\be
\left[ T^{ab}, Q^{c}\right]= -\frac{i}{2}\left( \epsilon^{ac}Q^{b}+
 \epsilon^{bc}Q^{a} \right).
\ee
At last, the anticommutator of two fermionic generators is given by~\footnote{
We use the following convention for the skew-symmetric tensor $\epsilon$:
$\;\epsilon_{ij} \epsilon^{jk}=\delta_i^k \;, \quad
\epsilon_{12} = \epsilon^{21} =1 \;$.}
\be
\left\{ Q^{Ai'i}, Q^{Bk'k}\right\} = -2\left(
 \alpha \epsilon^{AB}\epsilon^{i'k'}T_1^{ik}  +
  \epsilon^{ik}\epsilon^{i'k'}T_2^{AB}-
 (1+\alpha) \epsilon^{AB}\epsilon^{ik}T^{i'k'}
\right).
\ee
Here, $\alpha $ is an arbitrary real parameter. At $\alpha = 0$ and $\alpha = -1$ one of the
$su(2)$ algebras decouples and we recover the superalgebra $su(1,1\vert 2)\oplus su(2)$.
The superalgebra $D(2,1;1)$ is isomorphic to $osp(4^*\vert 2)$.
There are some different choices of the parameter $\alpha$ which lead to isomorphic
algebras $D(2,1;\alpha)$ \cite{FRS}.

For what follows it is convenient to pass to another notation,
\bea
&& P\equiv T_2^{22} \;, \quad K \equiv  T_2^{11} \;, \quad D\equiv -T_2^{12} \;, \nn
&& T\equiv T^{22} \;, \quad \bT\equiv  T^{11} \;, \quad T_3\equiv T^{12} \;, \qquad
 V\equiv T_1^{22} \;, \quad \bV\equiv  T_1^{11} \;,\quad V_3 \equiv  T_1^{12} \;,\nn
&&Q^i \equiv -Q^{21i}\;,\quad \bQ^i\equiv -Q^{22i}\;,\quad S^i\equiv Q^{11i}\; ,\quad
\bS^i\equiv Q^{12i}\;. \label{newgener}
\eea
One may check that $P$ and $Q^i, \bQ^i$ constitute the $N=4, d=1$ Poincar\'e superalgabra.
The generators $D, K$ and  $S^i, \bS^i$ stand for $d=1$ dilatations, special conformal
transformations and conformal supersymmetry, respectively.

We shall construct a nonlinear realization  of the superconformal group $D(2,1;\alpha)$
on the coset superspace parametrized as
\be\label{coset}
g=e^{itP}e^{\theta_i Q^i+\bt^i \bQ_i}e^{\psi_i S^i+\bpsi^i \bS_i}
e^{izK}e^{iuD}e^{i\phi V+ i\bphi \bV}~.
\ee
The coordinates $t, \theta_i, \bt^i$ parametrize the $N=4, d=1$ superspace. All other
supercoset parameters are Goldstone $N=4$ superfields. The stability subgroup
contains a $U(1)$ subgroup of the group $SU(2)$ realized on the doublet indices $i$, hence
the Goldstone superfields $\phi, \bar\phi$ parametrize the coset $SU(2)/U(1)$. Another
$SU(2)$ is entirely placed in the stability subgroup and acts only on
fermionic Goldstone superfields and $\theta$'s, mixing them with their conjugates.
With our choice of the $SU(2)$ coset we are led to assume that $\alpha \neq 0$. We could
equivalently choose another $SU(2)$ to be nonlinearly realized and the first one to belong
to the stability subgroup, then the restriction $\alpha \neq -1$ would be imposed
instead.

The left-covariant Cartan one-form $\Omega$ with values in the superalgebra $D(2,1;\alpha)$
is defined by the standard relation
\be
g^{-1}\,d\,g = \Omega~.
\ee
In what follows we shall need to know the explicit structure of several
important one-forms in the expansion of $\Omega$ over the generators \p{newgener},
\bea\label{cforms}
&&\omega_P=ie^{-u}d{\tilde t}\equiv e^{-u}\left[ dt+i\left( \theta_i d\bt^i +\bt^i d\theta_i
  \right) \right], \nn
&& \omega^Q_1=\frac{1}{\sqrt{ 1{+}\Lambda\bLam}}\left[ {\hat\omega}^Q_1 +
  \bLam {\hat\omega}^Q_2 \right], \;
\omega^Q_2=\frac{1}{\sqrt{ 1{+}\Lambda\bLam}}\left[ {\hat\omega}^Q_2 -
  \Lambda {\hat\omega}^Q_1 \right],  \nn
&& \bomega_Q^1=\frac{1}{\sqrt{ 1{+}\Lambda\bLam}}\left[ {\hat\bomega}_Q^1 +
  \Lambda {\hat\bomega}_Q^2 \right], \;
\bomega_Q^2=\frac{1}{\sqrt{ 1{+}\Lambda\bLam}}\left[ {\hat\bomega}_Q^2 -
  \bLam {\hat\bomega}_Q^1 \right],  \nn
&& \omega_D= idu-2\left( \bpsi^i d\theta_i + \psi_i d\bt^i \right)
  -2iz d{\tilde t} \; , \nn
&& \omega_V=\frac{1}{ 1{+}\Lambda\bLam}\left[ id\Lambda+{\hat \omega}_V+
 \Lambda^2{\hat\bomega}_{V}-\Lambda{\hat\omega}_{V_3}\right], \;
 \bomega_{V}=\frac{1}{ 1{+}\Lambda\bLam}\left[ id\bLam+{\hat \bomega}_{V}+
 \bLam^2{\hat\omega}_{V}+\bLam{\hat\omega}_{V_3}\right],\nn
&& \omega_{V_3}=\frac{1}{ 1{+}\Lambda\bLam}\left[ i\left( \Lambda d\bLam-\bLam d\Lambda\right)
+\left( 1{-}\Lambda\bLam\right){\hat\omega}_{V_3}
   -2\left( \Lambda{\hat\bomega}_{V}-\bLam{\hat\omega}_V \right)\right].
\eea
Here
\bea\label{cforms1}
&&  {\hat\omega}^Q_i=e^{-\frac{1}{2}u}\left( d\theta_i -\psi_i d{\tilde t}\right),\;
  {\hat\bomega}_Q^i=e^{-\frac{1}{2}u}\left( d\bt^i -\bpsi^i d{\tilde t}\right),\nn
&& {\hat\omega}_V=2\alpha \left[ \psi_2 d\bt^1 -\bpsi^1\left( d\theta_2-
       \psi_2 d{\tilde t}\right)\right], \;
 {\hat\bomega}_{V}=2\alpha \left[ \bpsi^2 d\theta_1 -\psi_1
    \left( d\bt^2-\bpsi^2 d{\tilde t}\right)\right],\nn
&& {\hat\omega}_{V_3}=2\alpha\left[ \psi_1 d\bt^1 -\bpsi^1 d \theta_1-
   \psi_2 d\bt^2 +\bpsi^2 d \theta_2 +
 \left( \bpsi^1 \psi_1 -\bpsi^2 \psi_2\right)d{\tilde t}\right],
\eea
and
\be \label{Lambda}
\Lambda = \frac{ \tan \sqrt{\phi\bphi}}{\sqrt{\phi\bphi}}\phi \; ,\;
\bLam = \frac{ \tan \sqrt{\phi\bphi}}{\sqrt{\phi\bphi}}\bphi \; .
\ee

The semi-covariant (fully covariant only under Poincar\'e supersymmetry)
spinor derivatives are defined by
\be
D^i=\frac{\partial}{\partial\theta_i}+i\bt^i \partial_t\; , \;
\bD_i=\frac{\partial}{\partial\bt^i}+i\theta_i \partial_t\; , \;
\left\{ D^i, \bD_j\right\}= 2i \delta^i_j \partial_t \; .
\ee

As the next step, we impose the inverse Higgs constraints \cite{IH}
\be\label{ih}
\omega_D=0\; , \quad \left. \left . \omega_V\right| = \bomega_V\right|=0 \; ,
\ee
where $|$ means spinor projections. These constraints are manifestly covariant
under the whole supergroup. They allow one to eliminate the Goldstone spinor superfields
and the superfield $z$ as the spinor and $t$-derivatives, respectively,
of the residual bosonic Goldstone superfields $u, \Lambda, \bar\Lambda$ and simultaneously imply
some irreducibility constraints for the latter:
\bea
&& i D^i u = -2 \bpsi^i\; , \; i\bD_i u=-2 \psi_i \; , \; \dot{u}=2z\; ,\nn
&& i D^1 \Lambda = 2\alpha\Lambda\left( \bpsi^1 + \Lambda \bpsi^2\right), \;
 i D^2 \Lambda = -2\alpha\left( \bpsi^1 + \Lambda \bpsi^2\right),  \nn
&& i \bD_1 \Lambda = 2\alpha\left( \psi_2 - \Lambda \psi_1\right), \;
 i \bD_2 \Lambda = 2\alpha\Lambda\left( \psi_2 - \Lambda \psi_1\right).
\eea

To understand the meaning of these constraints, let us pass to  the new variables
\be\label{nv}
q=e^{\alpha u}\frac{1-\Lambda\bLam}{1+\Lambda\bLam}\; ,\;
\lambda=e^{\alpha u}\frac{\Lambda}{1+\Lambda\bLam}\; , \;
\blam= e^{\alpha u}\frac{\bLam}{1+\Lambda\bLam}\; .
\ee
In these variables the constraints are rewritten as
\bea\label{ih2}
&& D^1\lambda = \bD^1   \lambda= 0 \; , \; D^2 \blam =\bD^2 \blam=0 \; , \nn
&& D^1 q=D^2 \lambda\; ,\; D^2 q=-D^1 \blam\; ,\; \bD_1 q=\bD_2 \blam\; ,\;
 \bD_2 q=- \bD_1 \lambda\; .
\eea
After introducing a new $N=4$ vector superfield $V^{ij}$ subject to
$V^{ij}=V^{ji}$ and $\overline{V^{ik}}=\epsilon_{ii'}\epsilon_{kk'}V^{i'k'}$ via
\bea\label{V}
&& V^{11}=-i\sqrt{2}\,\lambda \; ,\; V^{22}= i\sqrt{2}\,\blam\; , \; V^{12}=  \frac{i}{\sqrt{2}}\,q \;, \nonumber \\
&& V^2 \equiv V^{ik}V_{ik} = q^2 + 4\lambda\bar\lambda~,
\eea
the constraints \p{ih2} can be brought in the manifestly $SU(2)$-symmetric suggestive form
\be\label{tensor}
D^{(i}V^{jk)} =0 \; , \quad \bD{}^{(i}V^{jk)} =0 \; .
\ee
The superfield $V^{ik}$ subject to \p{tensor} is recognized as the one employed in \cite{ismi} for constructing
a general off-shell $N=4$ supersymmetric $1d$ sigma model with bosonic target manifold
of dimension $3k$.\footnote{The same off-shell $N=4, d=1$ supermultiplet was independently
considered  in \cite{bepa} and later rediscovered in \cite{stro2}.}
The constraints \p{tensor} leave in $V^{ik}$ the following independent superfield projections:
\bea
&& V^{ik}~, \; D^{i}V^{kl} = -\frac{1}{3}(\epsilon^{ik}\chi^l + \epsilon^{il} \chi^k)~, \;
\bar D^{i}V^{kl} = \frac{1}{3}(\epsilon^{ik}\bar\chi^l + \epsilon^{il} \bar\chi^k)~, \;
D^i\bar D^k V_{ik}~, \label{indep}
\eea
where
\be
\chi^k \equiv D^iV_i^k~, \quad \bar\chi_k = \overline{\chi^k} = \bar D_i V^i_k~.
\ee
Due to \p{tensor},
all other superfield projections are vanishing  or are expressed as time derivatives of \p{indep},
e.g.~\footnote{We use the notation:
$(D)^2 = D^iD_i~, \;(\bar D)^2 = \bar D_i\bar D^i~, \;[D,\bar D] = [D^i,\bar D_i]\,.$}
\be
(D)^2 V^{ik} = (\bar D)^2 V^{ik} = [D, \bar D] V^{ik} =0~, \quad D^{(i}\bar \chi^{k)} = \bar D^{(i}\chi^{k)} = 3i\dot V^{ik}~.
\ee
The spinor superfields $\chi^i$, $\bar\chi^k$ are related to the original Goldstone
fermionic superfields $\psi^i, \bar\psi^k$ by
\be
\chi^i = 2i\alpha V^{ik}\bar \psi_k~, \quad \bar\chi^i = -2i\alpha V^{ik}\psi_k~.
\ee

Note that the treatment of $V^{ik}$ in the $N=4$ superconformal mechanics context is somewhat
different from the one adopted in \cite{ismi,bepa}. In our case the superfield $V^{ik}$ provides an example of the construction
of a linear representation of $SU(2)$ symmetry in terms of its nonlinear realization.
As is seen from the identification \p{nv}, \p{V}, it is entirely of
Goldstone nature: its angular part $V^{ik}/|V|$ is related by an equivalence transformation to the $SU(2)/U(1)$
Goldstone superfields $\Lambda, \bar\Lambda$ while the norm
$|V| = \sqrt{q^2 +4\bar\lambda\lambda} = 1 + 2\alpha\,u + \cdots$ is related to the
dilaton $u$ and is non-vanishing for vanishing fields.

Important corollaries of eqs. \p{tensor} are the relations
\be\label{prop1}
(D)^2 \left( V^2\right)^{-\frac{1}{2}}=(\bD){}^2 \left( V^2\right)^{-\frac{1}{2}} =
[D, \bD{}] \left( V^2\right)^{-\frac{1}{2}}= 0
\ee
and
\be\label{prop2}
D_i \,\frac{ V^{ij} }{ \left( V^2\right)^{\frac{3}{2}} } =
\bD_i\, \frac{ V^{ij} }{ \left( V^2\right)^{\frac{3}{2}} } =0 \;.
\ee

The constraints \p{tensor} can be solved in terms of an unconstrained prepotential $\cW$,
\be\label{prepot}
V^{ij}= D^{(i}\bD^{j)}\cW~, \quad \bar \cW = -\cW~,
\ee
which is defined up to a gauge freedom
\be
\cW{\;}' = \cW + (D)^2\omega - (\bar D)^2\bar\omega  + [D,\bar D] \sigma, \label{prepgauge}
\ee
where $\omega $, $\bar \omega $ and $\sigma = -\bar\sigma$ are superfield gauge
parameters.

Let us quote the transformation properties of the $N=4$ superspace coordinates
and the basic Goldstone superfields under the transformations of conformal supersymmetry.
They are generated by acting on the coset element \p{coset} from the left by the element
\be
g_0= e^{\epsilon_i S^i +\beps^i \bS_i} \in D(2,1;\alpha)~.
\ee
Hereafter, when summing over doublet indices we assume them to stay in a natural position;
the Grassmann coordinates and their conjugates carry lower case and upper case indices, respectively.
We use as short-hand notation
$\bar\psi \cdot \xi = \bar\psi^i \xi_i = - \xi\cdot \bar\psi$.
The contraction of spinors of equal kind is defined as $a\cdot b = a^ib_i$,
$\bar a\cdot \bar b = \bar a_i\bar b^i$.

The sought transformations are
\bea
&& \delta t=-it \left( \epsilon \cdot\bt +\beps\cdot\theta \right) +(1+2\alpha)
  \theta\cdot\bt \left( \epsilon\cdot\bt -\beps\cdot\theta \right),\nn
&& \delta \theta_i= \epsilon_i t -2i \alpha \theta_i (\theta\cdot \beps) + 2i
  (1+\alpha) \theta_i (\bt\cdot\epsilon) -i (1+2\alpha) \epsilon_i (\theta\cdot\bt) \;, \nn
&& \delta u= -2i \left( \epsilon \cdot\bt +\beps\cdot\theta\right), \nn
&& \delta \Lambda= 2i\alpha\left[\theta_2\beps^1-\bt^1\epsilon_2+
\left(\bt^2\epsilon_1-\theta_1\beps^2\right)\Lambda^2+\left(\bt^1\epsilon_1-
  \theta_1\beps^1+\theta_2\beps^2-\bt^2\epsilon_2\right)\Lambda\right].
\eea
Under these, the $N=4$ superspace integration measure $dtd^4\theta$ is transformed
as
\be
\delta\; dtd^4\theta = - 2i(\epsilon\cdot \bt + \bar\epsilon\cdot \theta)\,dtd^4\theta~.
\label{meastrans}
\ee
The transformations of $V^{ij}$ defined in \p{V} can be written in the following manifestly covariant
fashion,
\be
\delta V^{ij} = -2i\alpha\left[ (\epsilon\cdot\bt+\beps\cdot\theta)V^{ij} +
(\epsilon^{(i}\bt_k-\beps_k\theta^{(i})V^{j)k}+
(\epsilon_k\bt^{(i}-\beps^{(i}\theta_k)V^{j)k} \right].\label{Vtrans}
\ee
As a consequence,
\be
\delta V^2 = -4i\alpha\,(\epsilon\cdot\bt+\beps\cdot\theta)\,V^2~. \label{Vtrans2}
\ee
The covariant derivatives $D^i$, $\bar D_i$ transform as
\bea
\delta D^i = i\left[(2 +\alpha)(\epsilon\cdot\bt) +\alpha (\theta\cdot \beps)\right]D^i
- 2i(1+\alpha)(\beps\cdot \bt)\bar D^i
 -2i\alpha \left[\epsilon^{(i}\bt_{k)} + \theta^{(i}\beps_{k)}\right] D^k, \label{Dtrans}
\eea
\bea
\delta \bar D_i = i\left[(2 +\alpha)(\beps\cdot\theta) +\alpha (\bt\cdot \epsilon)\right]\bar D_i
- 2i(1+\alpha)(\theta\cdot \epsilon)D_i
-2i\alpha \left[\epsilon_{(i}\bt_{k)} + \theta_{(i}\beps_{k)}\right]\bar D^k. \label{barDtrans}
\eea
{}From these transformations it follows, in particular, that chiral $N=4, d=1$ superfields exist
only at $\alpha = -1$, i.e. in the case of the supergroup $SU(1,1\vert 2)$.

The prepotential $\cW$ can be shown to have the following simple transformation rule,
\be
\delta \cW = -2i(1+\alpha)\,(\epsilon\cdot \bt + \beps\cdot \theta)\,\cW~. \label{cVtrans}
\ee
It is straightforward to check that the transformation laws \p{Dtrans}, \p{barDtrans} and
\p{cVtrans} for any $\alpha \neq 0$ imply for $V^{ik}$ in \p{prepot} precisely
the transformation \p{Vtrans}.

For completeness, we also give the variations of the $N=4, d=1$ superspace coordinates
under the $N=4, d=1$ Poincar\'e supergroup,
\be
\delta t = i\left(\theta\cdot\bar\varepsilon -\varepsilon\cdot\bar\theta\right)~, \quad
\delta \theta_i = \varepsilon_i~, \;\;\delta\bar\theta^i = \bar\varepsilon^i~.
\ee
All our superfields are scalars under the latter transformations.
Since all other $D(2,1;\alpha)$ transformations appear in the anticommutator of the conformal
and the Poincar\'e supersymmetry generators, it is sufficient to require invariance under these
two supersymmetries, when constructing invariant actions for the considered system.

\setcounter{equation}0
\section{Invariant actions in N=4 and N=2 superspaces}
We shall construct invariant actions for our variant of $N=4$
superconformal mechanics both in $N=4$ and $N=2$ superspaces.

Due to the basic constraints \p{ih2}, derivatives of each $N=4$ superfield with respect to, say,
$\theta_1 \equiv \xi, \bt^1 = \bar\xi$
can be expressed as derivatives with respect to $\theta_2 \equiv \theta,\bt^2 =\bar \theta$ of
other superfields. Therefore, only the $\xi=\bar \xi=0$ components of each $N=4$ superfield are
independent $N=2$ superfields. Let us denote these independent superfields as
\be\label{n2sf}
q|=v\;,\quad \lambda|=\rho \; ,\quad \blam|=\brho \;, \quad \cD \bar\rho = \cbD \rho =0~,
\ee
where $|$ means restriction to $\xi=\bar \xi=0$, and $\cD$, $\cbD$ are spinor derivatives with respect to
$\theta \equiv \theta_2$ and $\bt \equiv\bt^2$:
\be
\cD=\frac{\partial}{\partial\theta}+i\bt \partial_t\; , \;
\cbD = \frac{\partial}{\partial\bt}+i\theta \partial_t\; , \;
\left\{ \cD, \cbD \right\}= 2i \partial_t \; .
\ee

The transformations of the implicit $N=2$ Poincar\'e supersymmetry completing the explicit one to
the full $N=4$ have the following form in terms of these $N=2$ superfields:
\be
\delta v = - \varepsilon \cbD \brho -\bar\varepsilon \cD \rho\; , \quad
\delta \rho= \varepsilon\cbD v \;, \quad \delta\brho=\bar\varepsilon \cD v\;. \label{pt2}
\ee
The transformations of the two $R$-symmetry $SU(2)$ groups are realized as
\be\label{pt6}
\begin{array}{ccc}
\mbox{ $T,\bT$ transformations} & & \mbox{ $V,\bV$ transformations} \\
\left\{ \begin{array}{l}
\delta v = -a \theta \cbD\brho-\bar a \bar\theta \cD\rho\;,\\
\delta\rho=a\theta \cbD v\;, \\
\delta\brho=\bar a \bar\theta\cD v\;,
\end{array} \right.
&&
\left\{ \begin{array}{l}
\delta v = -2(b \brho+\bar b \rho)+b \bar\theta \cbD\brho+\bar b \theta \cD\rho\;,\\
\delta\rho=b v -b \bar\theta \cbD v\;, \\
\delta\brho=\bar b v - \bar b \theta\cD v\;,
\end{array} \right.
\end{array}
\ee
while the conformal supersymmetry $S^2,\bS_2$ transformations act as
\bea\label{pt7}
&& \delta v =2i\alpha \left(\theta\beps+\bar\theta\epsilon\right)-
 \epsilon\left( t+i\theta\bar\theta\right)\cD v -
 \beps\left( t-i\theta\bar\theta\right)\cbD v+ 2it\left(\epsilon\bar\theta+
  \beps\theta\right) \dot v \;,\nn
&& \delta \rho=4i\alpha\theta\beps \rho-\epsilon\left( t+i\theta\bar\theta\right)
   \cD \rho + 2it\left(\epsilon\bar\theta+ \beps\theta\right) \dot\rho \;, \nn
&& \delta \brho=4i\alpha\bar\theta\epsilon \brho-\beps\left( t-i\theta\bar\theta\right)
   \cbD \brho + 2it\left(\epsilon\bar\theta+ \beps\theta\right) \dot\brho \;.
\eea
The transformations \p{pt2}, \p{pt6}, \p{pt7} together with the manifest $N=2$ supersymmetry close
on $D(2,1;\alpha)$. The $S^1, \bar S_1$ transformations can be restored by commuting \p{pt7}
with \p{pt6}.

Invariant superfield actions consist of a superfield kinetic term and a superpotential.
The superfield kinetic term can be easily found to be
\be\label{free1}
S_1=\int dt d^4 \theta \left( V^2\right)^{\frac{1}{2\alpha}}=
 \frac{1+\alpha}{\alpha^2}\int dt d^2\theta
\left( v^2+4\rho\brho\right)^{\frac{1-2\alpha}{2\alpha}}
 \left( \cD v\cbD v +\cD \rho \cbD\brho\right),
\ee
where the $N=4$ superspace integration measure is related to the $N=2$ superspace one by
\be
d^4\theta = d^2\theta d^2 \xi~, \quad \int d^2\theta\, (\theta\bt) =1~, \quad
\int d^2\xi\, (\xi\bar \xi)  = 1~.
\ee
The invariance of \p{free1} immediately follows from the transformation properties of
the integration measure and of $V^2\,$, eqs. \p{meastrans}, \p{Vtrans2}.

The action \p{free1} vanishes for $\alpha=-1$, i.e. for the $SU(1,1|2)$ case,
as a consequence of \p{prop1}. Hence, in this situation we consider instead
\be\label{free1a}
S_{\alpha=-1}=
-\sfrac{1}{2}\int dt d^4 \theta \left( V^2\right)^{-\frac{1}{2}} \ln V^2=
\int dt d^2\theta \left( v^2+4\rho\brho\right)^{-\frac{3}{2}}
 \left( \cD v\cbD v +\cD \rho \cbD\brho\right).
\ee
Under $N=4$ superconformal transformations
the Lagrangian density in \p{free1a} changes
by a total derivative, in contrast to \p{free1} where the Lagrangian is a tensorial density.
The invariance of \p{free1a} immediately follows from the results of \cite{ikl2} where
a similar action was considered for the basic superfield having the same transformation properties
as  $(V^2)^{-1/2}$ and being subject to constraints more general than \p{prop1}.
The field content of \cite{ikl2} was different however.

Let us note that $\frac{1}{1+\alpha}S_1$ is regular for any $\alpha$ and coincides
with $S_{\alpha=-1}$ for $\alpha=-1$. In manifestly $N=4$ covariant notation
the action \p{free1a} is recovered from \p{free1} in the following way. The latter
is invariant for any nonzero $\alpha$. Let us expand its Lagrangian density
around $\alpha=-1$:
\be
\left( V^2\right)^{\frac{1}{2\alpha}} =
\left( V^2\right)^{-\frac{1}{2}-\frac{\alpha+1}{2}+O((\alpha+1)^2)} =
\left( V^2\right)^{-\frac{1}{2}} -\frac{\alpha{+}1}{2}
\left( V^2\right)^{-\frac{1}{2}} \ln V^2 + O\bigl((\alpha{+}1)^2\bigr)~.
\ee
The first term does not contribute to the action in virtue of \p{prop1},
and so we have
\be
\frac{1}{\alpha+1}\left[\left( V^2\right)^{\frac{1}{2 \alpha}}
                - \left( V^2\right)^{-\frac{1}{2}}\right] =
-\sfrac{1}{2} \left( V^2\right)^{-\frac{1}{2}} \ln V^2 + O(\alpha{+}1)~,
\ee
which yields just \p{free1a} in the limit $\alpha = -1$.

The potential term is easier to present in the $N=2$ superfield formulation:
\be\label{pt0}
S_p=
 \int dt d^2\theta\,  \ln\left( \frac{v+\sqrt{v^2+4\rho\brho}}{2}\right).
\ee
To construct this term, one starts from the obvious ansatz
\be\label{pt1}
S_p^{(0)}= \ln v = u + const
\ee
to which $S_p$ should simplify upon reduction to the $N=2$ superconformal mechanics based on
the supergroup $SU(1,1|1)$ \cite{ikl2}.
The variation of \p{pt1} with respect to \p{pt2} reads
\be\label{pt3}
\delta S_p^{(0)} = -\frac{\delta \left(\rho\brho\right)}{v^2} \;.
\ee
Thus to have an $N=4$ supersymmetric potential term we have to add a term which
compensates for \p{pt3}, namely
\be\label{pt4}
S_p^{(1)}= \frac{\rho\brho}{v^2} \;.
\ee
Iterating this recursive procedure, we eventually uncover the final form \p{pt0}
of the $N=4$ invariant potential term via
\bea\label{pt5}
&&\ln v + \frac{\rho\brho}{v^2} -\frac{3}{2}\left(  \frac{\rho\brho}{v^2}\right)^2 +
\frac{10}{3}\left(  \frac{\rho\brho}{v^2}\right)^3-
\frac{35}{4}\left(  \frac{\rho\brho}{v^2}\right)^4+\ldots \nn
&& = \ln v +\sum_{n=0}^{\infty}
(-1)^n \frac{ (2n+1)!}{(n+1)!(n+1)!} \left( \frac{\rho\brho}{v^2}\right)^{n+1}=
 \ln \left( \frac{v+\sqrt{v^2+4\rho\brho}}{2}\right).
\eea
One may check that this expression is invariant, up to a total derivative,
with respect to both types of $SU(2)$ transformations \p{pt6} as well as
with respect to the $S^2, \bar S_2$ transformations \p{pt7}.
Hence, it is $D(2,1;\alpha)$ invariant.

It is interesting to rewrite the potential term in a manifestly $N=4$ supersymmetric form, i.e.
in $N=4$ superspace in terms of the superfield $V^{ik}$. It turns out to be a superconformally
invariant version of the standard Fayet-Iliopoulos term $\sim\int dtd^4\theta \cW$:
\be
S_p = -i\,\sqrt{2}\,\int dtd^4\theta \left(\int^1_0 dy\, \partial_y \,\cW  \frac{1}{\sqrt{V^2}}\right). \label{n4pot}
\ee
The scale invariance of $S_p$ is easily seen.
Indeed, the integration measure has the
dilatation weight $(-1)$ (in units of length), while $V^{ik}$, $V^2$ and $\cW$ have weights
${\alpha } $, $2\alpha $ and $\alpha{+}1$, respectively. Then, the weight of the Lagrangian
density in \p{n4pot} is just $+1$, and \p{n4pot} is scale invariant. It is also easy to show its
invariance under the whole $N=4$ superconformal symmetry. One should take into account the
transformation
law of the superspace integration measure \p{meastrans} and the transformation laws \p{Vtrans2}
and \p{cVtrans}.
The invariance of $S_p$ under the gauge transformations \p{prepgauge} of the prepotential $\cW$ immediately
follows from the constraints \p{prop1}.
The extra parameter $y$ in \p{n4pot} defines a deformation $\cW(y)$ with
$\cW(1)=\cW$ and $\cW(0)=const$. Hence, the expression \p{n4pot} is analogous to
the standard representation of WZW terms on group manifolds via an integral over
an extra parameter \cite{pw}. In particular, we observe that in the variation
$\delta S_p$ this integral can be taken off
(one should use \p{prop2} while checking this):
\be
\delta S_p = -i\,\sqrt{2}\int dtd^4\theta\, \delta  \cW  \frac{1}{\sqrt{V^2}}~.
\ee
Therefore, this term makes a tensor contribution
$\sim 1/\sqrt{V^2}$ to the full $N=4$ superfield equations
of motion as obtained by varying the total action with respect to the prepotential $\cW$. This
WZW structure is not accidental: in the bosonic sector the superpotential \p{n4pot} contains the $d=1$ WZW term on
the sphere $S^2 =SU(2)/U(1)$.

Re-expressing the superpotential \p{n4pot} in the $N=2$ superfield form \p{pt5} is
rather involved though straightforward. One should make use of the relation
\be
\cW = -\frac{1}{6}\,\theta_i\bar\theta_kV^{ik} + \frac{1}{12} D^i(\theta^2\bt^kV_{ik}) -
\frac{1}{12} D^i(\theta^2\bt^kV_{ik})~,
\ee
which is valid modulo gauge transformations from \p{prepgauge}. Substituting this representation
in \p{n4pot} and integrating by parts, one brings \p{n4pot} to the form
\be
S_p = \frac{i\,\sqrt{2}}{24}\int\!\!dtd^4\theta \theta^2\bt^2 \int_0^1\!\!dy
\left[3 D_i\bar D_k \frac{\partial_yV^{ik}}{\sqrt{V^2}} +
      \bar D^k \frac{D^i\partial_yV_{ik}}{\sqrt{V^2}}  -
      D^k \frac{\bar D_i\partial_yV^{ik}}{\sqrt{V^2}} \right].
\label{interm}
\ee
After this one inserts the relation
$$
\theta^2 \bt^2 = - 4 (\theta\bt)\,(\xi\bar\xi)~,
$$
integrates in \p{interm} with respect to $\xi, \bar\xi$, expresses all $D_\xi, \bar D_\xi$ derivatives
through the $N=2$ ones $\cD, \cbD$ using the relations \p{ih2}, and finally integrates
by parts with respect to $\cD, \cbD$ in order to remove the delta function $\theta\bt$.
Important relations in the course of this calculation are
\be
\frac{1}{\sqrt{v^2 +4\rho\bar \rho}}(2\rho \cD v - v \cD\rho) =
\cD\left[\frac{-2\rho}{\sqrt{v^2 +4\rho\bar \rho}(v +\sqrt{v^2 +4\rho\bar\rho})}\right]
\ee
and its complex conjugate. The integral with respect to $y$ can be taken off at the end,
finally confirming the relation
\be
S_p =  -i\,\sqrt{2} \int dtd^4\theta \left(\int^1_0 dy\, \partial_y \,\cW
\frac{1}{\sqrt{V^2}}\right) =
\int dt d^2\theta \ln \left(v +\sqrt{v^2 + 4\rho\bar\rho}\right). \label{finPot}
\ee

\setcounter{equation}{0}
\section{The bosonic sector}

The bosonic worldline action, with all fermions discarded and the auxiliary field $[\cD,\cbD]v|$ eliminated by
its equation of motion, has the following form,
\bea\label{bosaction}
S_B &=& \mu^{-1}\frac{\alpha^2}{1+\alpha}\left(S_1\right)_B+\nu\left( S_p\right)_B \nn
&=& \int dt\biggl[\mu^{-1}\alpha^2\, e^u\,{\dot u}^2 +
 4\,\mu^{-1} e^u \frac{{\dot \Lambda}{\dot\bLam}}{(1+\Lambda\bLam)^2}
 -\frac{1}{4}\mu\, \nu^2e^{-u}
 + i\,\nu\frac{\bLam{\dot\Lambda}-\Lambda{\dot\bLam}}{1+\Lambda\bLam} \biggr]\;,
\eea
where $\mu$ is a constant of dimension of mass and the parameter $\nu$ is dimensionless.
It describes the dynamics of a particle with coordinates
$u(t)$, $\Lambda(t)$, $\bLam(t)$ in a three-dimensional target space.
In \p{bosaction}, the first term is the kinetic term of the dilaton~$u$,
the second one is the action of
the $SU(2)/U(1) \sim S^2$ nonlinear sigma model, the third one is the standard
dilaton potential, and the fourth one is a WZW term on $S^2$.
The dilaton potential arises after eliminating the auxiliary field from the sum of
\p{free1} and \p{finPot}, while the WZW term comes solely from \p{finPot}. The strict
relation between  the dilaton potential and WZW term is required by $N=4$ supersymmetry.

An equivalent form of the action is achieved by expressing it through
standard spherical coordinates $r$, $\vartheta$, $\varphi$ via
\bea
&& \Lambda = \tan \frac{\vartheta}{2} e^{i\varphi}\;, \quad
e^{u/2} = \frac{1}{\sqrt{2}}\mu r~, \label{av} \\
&& S_B=
\frac{1}{2}\,\int dt \left[4\alpha^2 \mu\,\dot r^2 +
 \mu r^2\left({\dot\vartheta}^2+\sin^2 \vartheta\, {\dot\varphi}^2  \right)-\frac{\nu^2}{\mu\,r^2} +
2\nu \cos \vartheta\, {\dot\varphi}\right]. \label{bosactionav}
\eea

It is also instructive to rewrite \p{bosaction} in such a way that
the potential term has the form typical for $3$-dimensional conformal mechanics. One
denotes $\bf{Y}$ $=(v, \rho, \bar\rho)$ and defines the new Cartesian
coordinates as $X^i = \mu^{-1}\,Y^i |{\bf Y}|^{\frac{1-2\alpha}{2\alpha}}$.
With this, \p{bosaction} is rewritten as
\be
S_B = \int dt \left[\mu\, g_{ik}(X) \dot{X}^i\dot{X}^k - \frac{1}{4 \mu} \frac{\nu^2}{|\bf{X}|^2} +
2i\nu \frac{\epsilon^{3ik}X^i\dot{X}^k}{(X^3 +|\bf{X}|)|\bf{X}|}\right], \label{bosact3}
\ee
where
\be
g_{ik}(X) = \delta_{ik} + (4\alpha^2 -1)\frac{X_i X_k}{|\bf{X}|^2}~.
\ee
 From this representation it is clearly seen that only at $\alpha = \pm 1/2$ the kinetic and
potential terms of $\bf{X}$ coincide with those of the standard 3-dimensional conformal
mechanics. At any other nonzero value of $\alpha$ one gets a conformally invariant
nonlinear sigma model which is a particular case of the general $1d$ sigma model
with $3k$-dimensional target space firstly considered in \cite{ismi}.
The advantage of the form \p{bosact3} is
that $SU(2)$ is manifest in all terms except for the WZW one.

To see which kind of geometry the sigma model in \p{bosaction}, \p{bosact3}
reveals, let us look at the line element corresponding to \p{bosactionav}
(with $\mu =1$ for simplicity),
\be
ds^2 = 4\alpha^2 dr^2 + r^2 d\Omega^2~, \quad
r = \textstyle{\sqrt{2}}\,e^{u/2} = |\bf{X}|~, \label{dist}
\ee
where
\be
d\Omega^2 = d\vartheta^2 + \sin^2\vartheta\,d\varphi^2
\ee
is the invariant Killing form on the sphere $S^2$.
This metric belongs to a 3-dimensional cone $C(B)$ over the base manifold
$B=S^2$ of radius $\frac{1}{4\alpha^2}$
(see \cite{gibry} for the relevant definitions).
At $\alpha=\pm 1/2$ one recovers flat space $\mathbb{R}^3$
(a cone over $S^2$ of unit radius), while at any other $\alpha$
one ends up with a curved manifold.
It is easy to evaluate the corresponding Ricci tensor and scalar curvature as
\be
R_{\vartheta\vartheta}=\Bigl(\frac{1}{4\alpha^2}{-} 1\Bigr)~, \quad
R_{\varphi\varphi}=\Bigl(\frac{1}{4\alpha^2}{-} 1\Bigr)\,\sin^2\vartheta~,\quad
R = \Bigl(\frac{1}{4\alpha^2}{-} 1\Bigr)\,\frac{2}{r^2}~,
\ee
and all other components being zero.
Thus for $\alpha \neq \pm 1/2$ our manifold is neither Ricci-flat nor
of constant scalar curvature.
Note that such a conical geometry is typical for
the bosonic sectors of superconformal theories in diverse dimensions \cite{gibry}.
The metric in \p{dist}, in agreement with the general finding of \cite{ismi}, is
conformally flat, which can be shown by changing the radial variable as
$r = \hat{r}^{\frac{1}{2|\alpha|}}$.

Finally, let us discuss some properties of the $S^2$ WZW term in \p{bosaction}.
It is not manifestly $SU(2)$ invariant: under the $SU(2)/U(1)$ transformations of
$\Lambda, \bar\Lambda$,
\begin{equation}
\delta \Lambda = b + \bar b\, \Lambda^2~, \quad 
\delta \bar\Lambda = \bar b + b \,\bar\Lambda^2~, \label{transfL}
\end{equation}
it is shifted by a full time derivative
\begin{equation}
\delta\; \frac{\bar{\Lambda} \dot{\Lambda} -
\Lambda\dot{\bar{\Lambda}}}{1+\Lambda\bar{\Lambda}} = \bar b\,
\dot{\Lambda}
- b\, \dot{\bar{\Lambda}}~, \label{3}
\end{equation}
i.e. it is invariant up to an abelian gauge transformation as should be. 
As was observed for the first time in \cite{jack}, 
the WZW term is conformally invariant on its own.
Also typical for WZW terms \cite{wit2,pw},
the constant $\nu$ needs to be quantized,
\be
\nu \in \sfrac12\mathbb{Z} \label{quantiz}
\ee
(see \cite{plyu} and refs. therein).
Thus the coefficient in front of the superpotential
in our $N=4$ superconformal mechanics should
be quantized for topological reasons.  As a consequence, the dilaton mass
(the coefficient of the dilaton potential in \p{bosactionav})
is proportional to the square of an integer.\footnote{Some other arguments why
this parameter of the bosonic conformal mechanics should be quantized were
adduced in \cite{6}.} This should be contrasted with
the $N=4$ superconformal mechanics of ref. \cite{ikl2} where no WZW term
is present in the bosonic sector and, therefore, no restriction
on the strength of the potential term arises.

To further clarify the meaning of the WZW term, let us
introduce a constant unit 3-vector $C^i$, $|\bf{C}|$ $=1$, and rewrite the WZW
term in \p{bosact3} (with the factor $2i\nu$ detached) as
\begin{equation}
L_{WZW} = \frac{\epsilon^{ikl}\,C^k\,X^l}{\left[(\bf{X}\cdot \bf{C}) +|\bf{X}|\right]\,
|\bf{X}|}\,\dot{X}^i
\equiv {\bf A}\cdot \dot{{\bf X}}~.
\end{equation}
Choosing the frame $C^3 =1,\; C^1=C^2 =0$, one reproduces just the WZW term in \p{bosact3}.
Now one observes that this term is nothing but the coupling of
a non-relativistic particle $X^i(t)$ to the potential $A^i$ of
a Dirac magnetic monopole, with the singular Dirac string
oriented along $\bf{X}=-\bf{C}$. It is easy to calculate the
corresponding magnetic field strength
\begin{equation}
F^{kl} = \partial^kA^l - \partial^lA^k =
\epsilon^{klp}\,\frac{X^p}{|{\bf X}|^3}
\end{equation}
which is indeed independent of~$C^i$.
It is also easy to check that the dependence of the WZW term on
the unit vector $C^i$ is reduced to a full time derivative (it is a
worldline pull-back of a gauge transformation of $A^i$) and, hence, the
action does not depend on this parameter up to topological effects.
This follows from the relation
\begin{equation}
\epsilon^{pij}\,C^i\,\partial_C^j \left( L_{WZW} \right) =
-\partial_t\left[\frac{X^p + C^p|\bf{X}|}{(\bf{X}\cdot \bf{C}) + |{\bf X}|} \right].
\end{equation}
Thus eq. \p{quantiz} is just the Dirac quantization condition.

\setcounter{equation}{0}
\section{Special bosonic dynamics}
In this Section we put $\mu =1$ for simplicity.
In standard spherical coordinates $\br=(x^i)=(r,\vt,\vf)$
our bosonic Lagrangian
\bea
{\cal L}_B &=& \frac{1}{2}\bigl(
4\alpha^2\,\dot{r}^2 + r^2\,\dot{\vt}^2 + r^2\sin^2\vt\,\dot{\vf}^2 \bigr)
-\frac{\nu^2}{2r^2} + \nu\cos\vt\,\dot{\phi} \nn
&=& \frac{1}{2}\bigl( 4\alpha^2\,\dot{r}^2 + \bL^2 r^{-2} \bigr)
-A_0(\br) + A_i(\br)\,\dot{x}^i
\eea
describes an electrically charged particle in a spherically symmetric
external electromagnetic field given by
\be
E_i\ =\ \nu^2 \frac{x_i}{r^4} \qquad\textrm{and}\qquad
B_i\ =\ - \nu\,\frac{x_i}{r^3} \;.
\ee
Note that $A_0{=}\nu^2/2r^2$ is just a central scalar potential while
$(A_r{=}0,A_{\vt}{=}0,A_{\vf}{=} \nu \cos\vt)$
yields a Dirac monopole of magnetic charge $g{=}{-}\nu $.
The two potentials are related by $N{=}4$ supersymmetry; this will
give rise to special dynamics.

Let us analyze the particle motion in more detail.
The angular dynamics is independent of~$\alpha$.
Obviously conserved is the canonical momentum
\be
p_{\vf}\ =\ r^2\sin^2\vt\,\dot{\vf} + \nu\cos\vt\ =:\ j
\ee
because $\vf$ is a cyclic variable.
The $\vt$ equation of motion
\be
\partial_t (r^2 \dot{\vt})\ =\
\frac{(j{-}\nu\cos\vt)(j\cos\vt{-}\nu)}{r^2\sin^3\vt}
\ee
still permits the solution $\dot{\vt}=0$ while $\dot{\vf}\neq0$ provided that
\be
j\,\cos\vt\ =\ \nu \quad,\qquad\textrm{i.e.}\quad \cos\vt\ =\ \textrm{const}\,,
\ee
which places the motion on a (two-dimensional) cone with opening angle~$\vt$
around the 3-axis. This choice of coordinates simplifies
\be
\dot{\vf}\ =\ \frac{j}{r^2} \qquad\textrm{and}\qquad
l^2\ :=\ \bL^2\ =\ j^2\sin^2\vt =\ j^2-\nu^2\ =\ \textrm{const.}
\ee

The radial motion is governed by
\be
4\alpha^2\,\ddot{r}\ =\ \frac{j^2}{r^3} \qquad\Longrightarrow\qquad
E\ :=\ \frac12\,4\alpha^2\,\dot{r}^2 + \frac{j^2}{2r^2}\ =\ \textrm{const}\,,
\ee
which is readily solved by
\be
r(\vf)\ =\ \frac{r_0}{\cos\frac{\vf}{2\alpha}}\label{111}
\ee
where $r_0=j/\sqrt{2E}$ is the minimal distance to the origin.
The trajectories are unbounded; the polar scattering angle is
$\Delta\vf=2\alpha\pi$. As usual, the motion is fixed by two data:
$l$ (or $j$ or $\vt$) and $E$.

Without the central potential $A_0(r)$ the particles would follow geodesics
on the 2D cone inside the 3D cone; the magnetic field merely exerts the
constraining force. The repulsive force provided by $A_0(r)$, however, is
of a special kind because it gives rise to an enhanced (`dynamical') symmetry,
at least in the flat case~$\alpha^2{=}1/4$:
Rotational invariance implies (\`a la Noether) the conservation of a canonical
angular momentum vector~$\bJ$ which, due to the magnetic potential,
differs from the kinematical angular momentum $\bL=\br\times\dot{\br}$ via
\be
\bJ\ =\ \bL + \nu\frac{\br}{r} \qquad\Longrightarrow\qquad
\bJ^2\ =\ l^2+\nu^2\ =\ j^2 \;.
\ee
{}From $\frac{\br}{r}{\cdot}\bJ= \nu$ we infer again that the motion is
constrained to a cone around~$\bJ$,
and $\bL$ is precessing with angular velocity $j/r^2$.
The equation of motion
\be
\ddot{\br}\ =\ \nu^2\frac{\br}{r^4} - \nu\frac{\dot{\br}{\times}\br}{r^3}
\ =\ \nu\frac{\bJ}{r^3}
\ee
confirms, of course, that $\dot{\bJ}=\br\times(\textrm{equation of motion})=0$
irrespective of the central potential.
For $A_0{=}0$ this system has been analyzed, e.g. in \cite{plyu} where also 
hidden symmetries in the absence of an electric potential were discussed.
The special form of $A_0$, however, keeps the acceleration $\ddot{\br}$ in
a fixed direction ($\sim\bJ$) which implies
\be
\ddot{\br}\times\bJ\ =\ 0 \qquad\Longrightarrow\qquad
\dot{\br}\times\bJ\ =:\ \bI\ =\ \textrm{const.} \qquad\Longrightarrow\qquad
\dot{\br}\cdot\bI\ =\ 0
\ee
meaning that the trajectory is planar ($\perp\bI$).
This result was already apparent in \p{111} for $2\alpha{=}1$.
Hence, the particle follows a special hyperbola.
Yet another way to seeing this employs the effective potential
\be
V_{\textrm{eff}}\ =\ \frac{l^2}{2r^2}+\frac{\nu^2}{2r^2}
\ =\ \frac{j^2}{2r^2}
\ee
which, together with $r^2\dot{\vf}=j$, shows that the projection onto the
$\vt{=}\frac{\pi}{2}$ plane ($\perp\bJ$) gives just free motion, i.e.
a straight line, with $l$ replaced by~$j$.

To characterize the symmetry enhancement present for our particular
central potential~$A_0$ (in combination with the magnetic potential~$A_i$),
we compute the Poisson brackets of $\bJ{=}(J_k)$ and $\bI{=}(I_k)$
in flat space (for $\alpha^2=1/4$)~\footnote{
The magnetic field modifies the canonical brackets:
$\{p_k,p_l\}=\eps_{klm}B_m$ for the momenta $p_k{=}\dot{x}_k$.}
and find
\be
\{J_k,J_l\}\ =\ \eps_{klm} J_m \quad,\qquad
\{J_k,I_l\}\ =\ \eps_{klm} I_m \quad,\qquad
\{I_k,I_l\}\ =\ -2H\eps_{klm} J_m \label{dynsym}
\ee
where $H=\frac12({\bf p}^2+\frac{\nu^2}{r^2})$ denotes the Hamiltonian of
our system. It is part of the symmetry algebra and conserved with value $H=E$.
The relation $\bI^2=2H\bL^2$ may be exploited
to directly find the trajectory via
\be
\sqrt{2E}\,l\,r\sin\vt\cos\vf\ =\ |\bI|r\sin\vt\cos\vf\ =\ \br{\cdot}\bI\ =\
\br{\cdot}(\dot{\br}{\times}\bJ)\ =\ \bL{\cdot}\bJ\ =\ \bL^2\ =\ l^2
\ee
which yields $r(\vf)=\frac{l/\sin\vt}{\sqrt{2E}\cos\vf}$ once more.
The algebra \p{dynsym} is very reminiscent of the one generated in the Kepler
problem by the angular momentum and the Runge-Lenz vector, in the $E{>}0$
case of unbounded motion. Indeed, rescaling $I_k$ reveals
an $SO(3,1)$ symmetry algebra, just like in the latter case.

\vfill\eject

\setcounter{equation}{0}
\section{Concluding remarks}
In this paper we have presented a new variant of $N{=}4$ superconformal 
mechanics based on a nonlinear realization of the most general 
$d{=}1, N{=}4$ superconformal symmetry associated
with the supergroup $D(2,1;\alpha)$. In the bosonic sector our version 
contains on the worldline, besides the dilaton field, two fields 
parametrizing the two-sphere $S^2 \sim SU(2)/U(1)$.
This field content suggests that the model can be mapped, by a generalization 
of the equivalence transformation of~\cite{bik3,ikn}, onto 
$N{=}4$ AdS$_2\times S^2$ superconformal mechanics describing a charged 
AdS$_2\times S^2$ superparticle, and so may be a `disguised' form of the latter 
(perhaps, for a special value of $\alpha$).

The nonlinear realization superfield techniques allowed us to construct the
relevant action in a manifestly $N{=}4$ supersymmetric way in terms of off-shell
$N{=}4$ superfields subject to the constraints proposed in~\cite{ismi}. 
We also gave an $N{=}2$ superfield form of the action. There exist two separate
invariants which, respectively, extend the kinetic term of the dilaton combined 
with an $SU(2)/U(1)$ nonlinear sigma model as well as the potential term of 
dilaton. For the latter case, $N{=}4$ supersymmetry requires the dilaton 
potential to be accompanied by a $d{=}1$ WZW term on $S^2$. 
The former may be viewed as the potential of a radial electric field, 
and the latter is interpreted as the coupling to a Dirac magnetic monopole. 
The coupling constant in front of the WZW term is topologically quantized.
It simultaneously defines the dilaton mass, and so the latter
is also quantized in the $N{=}4$ superconformal mechanics under consideration.
The target space of the full bosonic sigma model for arbitrary nonzero values of
the parameter $\alpha$ turned out to be a cone over~$S^2$.
The relevance of such conical geometries to superconformal theories was
pointed out in \cite{gibry}.
Furthermore, we have studied the classical dynamics of the bosonic sector 
of the model.  Its three worldline fields are the coordinates of a 
nonrelativistic massive charged particle moving in the conical $3D$ manifold 
equipped with a certain electromagnetic background.
The particle motion is characterized by an interesting interplay between the
electric and magnetic forces which restrict the trajectory to the intersection
of a $2D$ cone and a plane. In fact, the special form of the potentials gives
rise to a dynamical $SO(3,1)$ symmetry involving the Hamiltonian, just like
for unbounded motion in the Kepler problem.

It is worth remarking on the relation to previous studies. In components and in 
$N{=}1$ superfields, the $N{=}4$ superconformal mechanics associated with 
$D(2,1;\alpha)$ was already discussed in \cite{papa1} as a special class of $1d$
supersymmetric sigma models \cite{papa2}. However, no detailed form of
$D(2,1;\alpha)$ invariant actions was presented.
The $D(2,1;\alpha)$ superconformal mechanics models
(actually, the $\alpha{=}0$ and $\alpha{=}{-}1$ special cases)
made their appearance also as tools for describing the near-horizon geometry
of four-dimensional multi black holes in \cite{stro1,stro2}.
Yet, no explicit examples of a superconformal $N{=}4$ superfield action and/or
an $N{=}4$ superpotential were given there.

Besides the problems of establishing the explicit relationship with
the AdS$_2\times S^2$ superparticle and of considering the quantization
of the presented system (constructing the relevant Hamiltonian etc.),
there are other directions in which the results of this paper
could be extended. First of all, it is interesting to
elaborate on alternative nonlinear realizations of $D(2,1;\alpha)$
which could give rise to as yet unknown versions of $N{=}4$ superconformal
mechanics. For instance, we might place all the generators of one $SU(2)$ in 
the coset, leaving the second $SU(2)$ in the stability subgroup, or else break
both $SU(2)$ groups by considering the coset $SU(2)\times SU(2)/SU(2)_{diag}$
or the coset $SU(2)/U(1) \otimes SU(2)/U(1)\sim S^2\otimes S^2$.
The corresponding versions of $N{=}4$ superconformal mechanics, if existing,
should contain three (in the first two cases) or four (in the third case)
physical bosonic fields besides the dilaton in their bosonic sectors.
It would be interesting to examine whether there exist suitably constrained
$N{=}4, d{=}1$ superfields capable of accommodating these field contents.
We expect that the precise form of the constraints is predicted by the
nonlinear realization formalism, like in the case considered in this paper.
In the four-field cases, $d{=}1$ versions of the hypermultiplet are likely
to be recovered in this way.

Another intriguing problem is the construction of $N{=}8$ superconformal
mechanics. It was recently noticed \cite{de} (see also \cite{zu,smi})
that one can realize off-shell $N{=}8, d{=}1$ Poincar\'e supersymmetry
on a $N{=}4$ chiral superfield and the $N{=}4$ superfield $V^{ik}$
subject to the constraints \p{tensor}.
This representation is a $d{=}1$ reduction of the vector $N{=}2, d{=}4$
multiplet, just like the $N{=}4$ multiplet described by $V^{ik}$ is a reduction
of the $N{=}1, d{=}4$ vector multiplet~\cite{ismi}.
The corresponding manifold of physical bosons is 5-dimensional.
This matches nicely with the bosonic content of the hypothetical $N{=}8$
conformal mechanics based on a nonlinear realization of one of possible
$N{=}8, d{=}1$ superconformal groups, viz. $OSp(4^*|4)$ having
$SU(2)\times SO(5)$ as $R$-symmetry \cite{vanpr} and containing the
$N{=}4, d{=}1$ superconformal group $SU(1,1|2)$ as a supersubgroup \cite{FRS}.
The relevant supercoset should include the dilaton and the four coordinates
of the 4-sphere $S^4 \sim SO(5)/SO(4)$ as the basic Goldstone superfields.

\section*{Acknowledgements}
O.L. would like to thank N.~Dragon for useful discussions on sect.~5.
The work of E.I. and S.K. was supported in part by
an INTAS grant, project No 00-00254, DFG grant, project No 436
RUS 113/669, RFBR-DFG grant, project No 02-02-04002, RFBR-CNRS grant,
project No 01-02-22005, and a grant of the Heisenberg-Landau program.
They are grateful to the Institute for Theoretical Physics in Hannover
for the warm hospitality extended to them during the course of this work.

\vfill\eject

\end{document}